%% file: main.tex
\documentclass[a4paper,12pt]{book}
\pdfoutput=1
\usepackage[T1]{fontenc}
\usepackage{subfiles}
\usepackage[english]{babel} % If you write in English
\usepackage{a4wide}
\usepackage{graphicx}
\graphicspath{{images/}}
\usepackage{subfig}
\newlength\figureheight
\newlength\figurewidth
\usepackage{ifthen}
\usepackage{ifpdf}
\ifpdf
\usepackage[pdftex]{hyperref}
\usepackage{lettrine}
\else
\usepackage{hyperref}
\fi
\usepackage{color}
\hypersetup{%
colorlinks=true,
linkcolor=black,
citecolor=black,
urlcolor=black}

\usepackage{fancyhdr}
\let\headruleORIG\headrule
\renewcommand{\headrule}{\color{black} \headruleORIG}

\usepackage{colortbl}
\arrayrulecolor{black}

\fancypagestyle{plain}{
  \fancyhead{}
  \fancyfoot[C]{\thepage}
  
}

\makeatletter
\def\@textbottom{\vskip \z@ \@plus 1pt}
\let\@texttop\relax
\makeatother

\makeatletter
\def\cleardoublepage{\clearpage\if@twoside \ifodd\c@page\else%
  \hbox{}%
  \thispagestyle{empty}%
  \newpage%
  \if@twocolumn\hbox{}\newpage\fi\fi\fi}
\makeatother
\usepackage{amsthm}
\usepackage{amssymb,amsmath}
\usepackage{array}
\usepackage{bm}
\usepackage[footnote]{acronym}
\usepackage[resetlabels,labeled]{multibib}
\usepackage{url}
\newcites{URL}{Netography}

\usepackage{longtable}
\usepackage{multirow}
\usepackage[most]{tcolorbox}
\usepackage{mdframed}
\usepackage{lipsum}
\usepackage{lscape}
\usepackage{listings}
\usepackage{lettrine}
\usepackage{float}
\usepackage{fancyhdr}
\usepackage{nccrules}
\usepackage{titlesec}
\usepackage{verbatim}
\usepackage{longtable}
\usepackage{float}
\usepackage{enumitem}
\usepackage{pifont}
\usepackage{mathtools}
\usepackage{amsmath}
\begin{document}

\subfile{pageDeGarde}
\include{Chapitres/introduction}
\include{Chapitres/chapitre1}
\include{Chapitres/chapitre2}
\include{Chapitres/chapitre3}
\include{Chapitres/chapitre4}
\include{Chapitres/chapitre5}

\include{Chapitres/conclusion}
\bibliographystyle{plain}
\bibliography{references}\addcontentsline{toc}{chapter}{Bibliography}
\bibliographystyleURL{plain}
\bibliographyURL{urls}\addcontentsline{toc}{chapter}{Netography}
\end{document}

%% file: pageDeGarde.tex
\begin{titlepage}
\begin{center}

\vspace*{0.8cm}
{\Large  {\textbf{ Automate migration to microservices architecture
using Machine Learning techniques}}}\\[0.7cm]

\begin{tabular}{*{2}{>{\centering}p{.5\textwidth}}}
\large Meryam Chaieb & \large Mohamed Aymen Saied \tabularnewline
Laval University & Laval University \tabularnewline
Québec, QC, Canada & Québec, QC, Canada \tabularnewline
\text{meryam.chaieb.1@ulaval.ca} & \text{mohamed-aymen.saied@ift.ulaval.ca} 
\end{tabular}
\\[3cm]

\textbf{\LARGE Abstract} \\
 \vspace*{1cm}
 
The microservice architectural style has many advantages such as scalability, reusability, and easy maintainability. Microservices have therefore become a popular architectural choice when developing new applications. Reaping these benefits requires redesigning the monolithic application and moving them
to a microservices-based architecture. This process is inherently complex and costly, so automating this task is critical. \\
This report proposes a method by which potential microservices can be identified from a given monolithic application while treating this problem as a clustering thematic.\\
Our method
takes as input the source code of the source application in order
to apply different approaches to devise the one box application into its different microservices.
In this report we detail each of these techniques while finishing with a discussion and comparison of the results.\\

\vspace*{0.5cm}
\textbf{Keywords:} Machine learning, Monolithic architecture, Microservices architecture, Static analysis of source code, NLP.\\
\end{center}
\end{titlepage}

%% file: Chapitres/introduction.tex
\chapter*{ Introduction}

\label{chap:intro}

\paragraph{}
Software engineering is the science that implements software products to solve real world problems or to facilitate certain tasks. In the last decades, as technology has invaded
everyday life, these software products are facing thousands of new users for applications
that were previously designed for a limited number of people.

Over the past decades, software engineering research identified and attempted to solve a variety of issues pertaining to several phases of the software lifecycle.
\newline  However, the fast pace of evolution in the IT industry  and the staggering growth of new technologies \cite{vayghan2021kubernetes} 
based on APIs \cite{saied2015could1, shatnawi2018identifying, mujahid2021toward}, containers \cite{vayghan2019kubernetes}, 
microservices \cite{sellami2022improving, almarimi2019web, sellami2022hierarchical, saidani2019towards}, 
cloud and virtualization, put an increasing pressure on software development \cite{benomar2015detection} and deployment \cite{vayghan2019microservice, vayghan2018deploying} 
practice to fully exploit this paradigm shift. This led to constant questioning of existing techniques \cite{saied2015could1} and results of software 
engineering research \cite{saied2020towards, saied2018improving}, leading to investigating the use of AI and ML-based techniques to solve software engineering problems 
in topics related to software reuse \cite{gallais2020api}, recommendation systems \cite{saied2016automated}, mining software repositories \cite{saied2020towards}, 
software data analytics and patterns mining \cite{saied2018towards, huppe2017mining, saied2016cooperative, saied2015mining} , 
program analysis and visualization \cite{saied2015observational, saied2015visualization}, testing in the cloud environment, Edge-Enabled systems \cite{mouine2022event}, 
microservices architecture \cite{sellami2022combining} and mobile applications.
 
We then observed an increased tendency of organizations to move existing enterprise
applications to the cloud, also known as cloud computing. This corresponds to the access
to IT services (servers, storage, networking, software) through the Internet provided by a
provider. There are many reasons for this migration, for example: high availability, automatic scaling, easier management of the infrastructure and compliance
with the latest security standards guarantee a more agile and combined development and
operation flow.
\paragraph{}
Driven by this new paradigm, the design, creation, deployment and maintenance of
enterprise applications have fundamentally changed. To bridge this gap and make existing
monolithic applications ready for injection into cloud technology, they must run as
flexible, loosely coupled compositions of specialized services, which has recently become
the microservices style of architecture.
\paragraph{}
In addition, monolithic applications that have been developed over the years can
become large, complex and even inefficient. Resulting in obscure structures. On the one
hand, this makes the monolith difficult to maintain with reasonable effort and to adapt to new and better technologies.
\newline On the other hand, they are often unable to scale at the module level, but rather by
duplicating instances of the entire application. In most cases, this is an ineffective
approach to meeting rapidly changing workloads while maintaining optimal resource
utilization.

This is why the software industry invests today in research in the area of migration from monolithic applications recognized as application encapsulated
in a single box to the new architectural style of microservices where the application is
divided into several boxes each representing a particular service. 
\newline Unlike monolithics, microservices applications meet the requirements of the cloud perfectly
and their maintenance and development efforts are known to be reasonable.
\paragraph{}
During this report, we tackle the problem of migrating monolithic applications to
the architecture of highly available microservices, taking advantage in this quest of the
evolution of research in terms of artificial intelligence. The approach we present in this
report helps experts by recommending a way to decompose and it is up to the expert to
decide, based on the results of the evaluation metrics and his experience, about the final
composition of his microservices.
\paragraph{}
The work carried out is presented in this report organized as follows: 
\newline We start with a first chapter that defines the general context of the project and the
different objectives to be achieved as well as the desired results, concluding by presenting
the methodology that we adopt for our work. We then move on to the state of the art where we present the
various key concepts and related works distilled from the literature and we finish by defining the evaluation metrics of the project. In the third chapter we detail the different
approaches we propose to solve this problem. The next chapter takes care of the definition
of the different algorithms developed and used to finally conclude with the realization
chapter where we expose our realization environment as well as the obtained results
ending with a discussion. We end this report with a conclusion while presenting our
contributions and our perspectives.

%% file: Chapitres/chapitre1.tex
\chapter{General overview
}
\label{chap:chap1}
\textbf{\Large Introduction}
\paragraph{}
We begin this first chapter of the report by defining the general context of the
project. We then describe the project at a high
level while detailing our objectives and expected results, and we finish by explaining the
project development methodology.

\section{Project context}
In this section, we define the problematic of our project, the objectives and the
expected results.
\subsection{Objectives }
\paragraph{}
Monolithic applications suffer from a dependency hell where everything is tightly
coupled, resulting in long integration times and an inability to track down the source of
errors detected during the integration cycle. The scalability of a single application is
limited and the system suffers severe downtime during upgrades.
\newline With increasing demands for scalability and frequent maintenance of software products
and IT services, companies are adopting a microservices architecture for most of their
products to realize many benefits.
\newline Such architectures offer technical flexibility, faster production processes, reusable
functionality and even multi-disciplinary development teams. 
\paragraph{}
For these reasons, many companies that previously had products built with a monolithic
design have chosen to move their services to a microservices-oriented architecture. However,
this has proven to be a time consuming and very expensive process. 
\newline The manual migration process forces developers to understand the legacy code base
without proper documentation and break it down into reusable microservices without
sacrificing the functionality of the original application. 
\paragraph{}
In this project, we are interested in solving the problems encountered by the
practitioner, identified previously, when decomposing the monolithic application into
microservices. Our tool will serve as a support for the user by proposing a way to
decompose the single box system into multiple services taking into account different
criteria that we detail in the following chapter.

\subsection{Expected Results}
At the end of the project, these requirements should be met :

\begin{itemize}[label=\textbullet, font=\LARGE ]
    \item  Collect data from monolithic applications.
    \item  Implement a prototype based on existing tools from state of the art approaches for
decomposing a monolithic application to microservices.
    \item Select a tool to evaluate the decomposition.
    \item Conduct experiments to evaluate the proposed approach in terms of the collected
monolithic applications.
    \item The code must be well documented and written in a simple and succinct manner.
    \item The results should be better than the existing results or at least close to them.
\end{itemize}
\section{Methodology of work}
\paragraph{}
After having defined the project objectives, project management technique must be
implemented to streamline the solution development cycle.
\newline To improve product quality and accelerate the project development process, we adopted the CRISP-DM methodology as our project development methodology which
is illustrated in Figure 1.1.\\
The choice of CRISP-DM as a working methodology comes down to the fact that it is an
adaptable process model offering an overview of the data mining life cycle (the sequence
of phases is not strictly established).
\newline Our work focused in a first part on data collection and
exploration. Then we focused on the modeling and evaluation of the solution. Finally we
return to the correction and exploitation of the data so that they represent our problematic
at best, which leads us to the desired results.

\begin{figure}[H]
    \centering
   \includegraphics[bb=0 0 9cm 9cm]{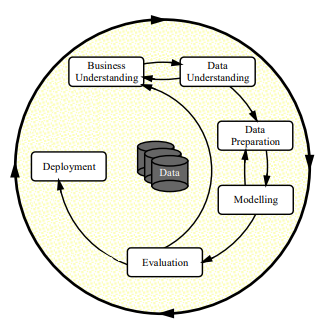}
    \caption{CRISP-DM Process Diagram} \cite{wirth_crisp-dm_nodate}
    \label{Fig:CRISP-DM Process Diagram}
\end{figure}
\paragraph{}
The CRoss Industry Standard Process for Data Mining (CRISP-DM) consists mainly of six
sequential phases as follows \cite{wirth_crisp-dm_nodate}:

\begin{itemize}[label=\textbullet, font=\LARGE ]
    \item \textbf{Business Understanding:} Business Understanding is the first phase of CRISP-DM
and focuses on determining the project objectives and requirements. In our case, this
step is detailed in the next chapter of the report entitled
"State of the Art" where we review key concepts and existing research that has
attempted to address similar problems.
    
    \item \textbf{Data Understanding:} Next comes the data understanding phase. Building on
the foundation of business understanding, it focuses on identifying, collecting and
analyzing the data sets that can
help to achieve the objectives of our project. During this stage, data from the
monolithic applications was collected using a static code analysis tool which will be
detailed in the following parts of the report. This phase is also illustrated in the third
chapter named "Proposed Approaches".
    
    \item \textbf{Data Preparation:} The necessary data sets are created in this phase. The data goes
through a selection, cleaning and reformatting pipeline. This step is also
described in the third chapter. 
    
    \item \textbf{Modeling:} At this stage, after choosing the modeling techniques to work with,
their parameters must be adjusted until the best
results are obtained. This phase is presented in the "Modeling" chapter. 
    
    \item \textbf{Evaluation:} In this critical step, the results of the models are evaluated to determine
the next step in the process. This phase represents an evaluation of the success of the
previous phase. In our report, this part is presented
during the last chapter named "Realization".
    
    \item \textbf{Deployment:} Beyond this phase the customer will have access to the software
product and to do so a monitoring and maintenance strategy must be created,
followed by a final report and a project review. In our case, the code must be clear
and concise since it
will be used in the future work of the laboratory's research team.

\end{itemize} 

\textbf{\Large Conclusion}
\paragraph{}
In this first chapter, we presented the general context of the project while detailing the objectives and the
expected results. Finally, we have proposed the development process of the project which
we adopt in the next chapters.

%% file: Chapitres/chapitre2.tex
\chapter{State of the art }
\label{chap:chap2}

\textbf{\Large Introduction} 
\paragraph{}
We begin this second chapter of the report by listing the different key concepts leading
to the good comprehensibility of the project. Then, we explain the PRISMA research
methodology that we used during the selection of scientific articles followed by our literature
review. We conclude this first part with a discussion mentioning our scientific
contributions to answer the first research question raised in this chapter.
\newline Finally, in a second part of the chapter, we answer a second research question by
presenting the different evaluation metrics distilled from the literature.

\section{Key concepts}

For a detailed description of our project, we start by defining the different key aspects
of our problem.

\subsection{ Software engineering}

Software engineering is the science of industrial engineering that studies the work
methodologies and best practices of engineers who develop software. Software engineering
applies engineering principles and techniques to the design of software systems such as the
definition of the architecture to be respected when creating a project.

In addition to the documentation tasks inherent to the follow-up of any engineering
project, the software engineer must be able to perform the tasks of analyzing the needs and
requirements of the software product, to design and develop it following professional
standards. Also, the software engineer ensures the necessary and adequate tests to guarantee the
conformity of its product to the previously defined specifications as he can still maintain his solution \citeURL{URL2}.
\paragraph{}
One of the famous architectures that has been the support of software product
development in the parent companies of the field for years is the monolithic architecture
that we define in the following section.

\subsection{Monolithique architecture}
\paragraph{}
Monolithic architecture was the traditional approach to software development,
previously used by large companies like Amazon and Ebay.
\newline In this architectural style, the functions are encapsulated in a single system. In other words,
it is a single box that gathers all the functionalities as illustrated by Figure 2.1.

Monolithic applications, as long as they are not complicated, have their own advantages
such as ease of use, development, testing and deployment.
\newline However, when the application tends to become more complicated, the monolithic
structure expands, becoming a large piece of software that is difficult to manage and scale.
The problem is that at present, many companies still have their software as monoliths. The
scalability and maintenance problems previously presented force companies to buy new
software instead of developing new features in the ones they have\cite{de_lauretis_monolithic_2019}.\\
Migrating their monolithic applications to microservices architecture can be a solution
and their software will regain its scalability and all the benefits associated with
microservices architecture.

\subsection{ Microservice architecture}
\paragraph{}
A new style of architecture called microservice, promises to address the problems of
monolithic architecture cited in the previous section. This is a trend in the practice of the
software engineering industry that has been defined by Lewis and Fowler \citeURL{URL3}.This
architectural style is considered a technique for developing software applications by
inheriting the principles and concepts of Service-Oriented Architecture (SOA). It allows
structuring an application based on services. 
\newline A service is a collection of small, loosely
coupled software services as illustrated in Figure 2.1. 
\newline Microservices can be seen as a new
paradigm for programming applications through the composition of small services, each
running its own processes and communicating via lightweight mechanisms. 

Microservices are very small entities, as their name indicates, for their contribution to
the application and not because of the number of lines of code.
\newline The main characteristics of microservices architectures are limited contexts, flexibility and
modularity. Indeed, each entity supports a single service, generally a single use case, and
the set of these microservices that constitute the application will thus ensure all the
services or functionalities guaranteed by the parent monolithic application. These loosely
coupled services can be developed independently.
\paragraph{}
In this regard, microservices are a new trend in software architecture that emphasizes
the development and design of highly maintainable and scalable software \cite{de_lauretis_monolithic_2019}.

\subsection{Refactoring  of monolithic applications}
\paragraph{}

Refactoring is an activity that aims to extend the life of existing software products. It is
then a transformation of the code while guaranteeing the same functionalities in order to
improve the source code that has structurally deteriorated with time or has accumulated
technical deficiencies.\\ 
A lot of research has been done in this area, mainly aiming at refactoring at the source
code level. Fowler and al  \cite{fowler_refactoring_nodate}  have consolidated the field through their well known books
entitled "Improving the design of Existing Code, more than 70 Patterns explained" where
they present refactoring techniques based on decomposition patterns. Dietrich in turn calls
code-level refactoring as high-impact refactoring \cite{dietrich_existence_2012}. 
\paragraph{}
The principle of migrating from a monolithic architecture to a more reliable
microservice architecture is to identify contextually related modules and encapsulate them
in a single service, while ensuring strong cohesion within a single microservice and weak
coupling between microservices.
\newline To get the most out of this refactoring step, choosing the right level of granularity when
dividing the functionality is a decisive step. However, building a new application from
scratch based on a microservices architecture can be a very expensive and time consuming task.
\newline On the other hand, the process of refactoring a mature monolithic application into
microservices also suffers from the same time and cost constraints. It is in this quest for
optimization that the industry today invests in research and development in the area of
refactoring and application migration \cite{bruel_monolith_2019}.

\begin{figure}[H]
    \centering
    \includegraphics[bb=0 0 15cm 8cm]{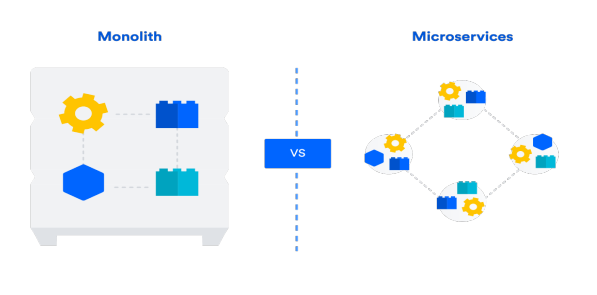}
    \caption{Monolithic application Vs Microservices application}\protect \citeURL{URL6}
    \label{Fig:- Monolithic application Vs Microservices application}
\end{figure}

\subsection{Machine learning}
\paragraph{}
Machine learning is a form of artificial intelligence, a field that combines computer
science and robust data sets to enable problem solving by aiming to copy the decision making process of a human being. The main goal of machine learning is to create systems
that learn and improve their performance by examining a stream of data to train and learn
models, and then produce descriptions and predictions that ensure optimal decision
making, even in the most complex scenarios.\citeURL{URL5}

\subsection{ Natural language processing }
\paragraph{}
 Natural language processing  is one of the most dynamic research areas in
the field of data science. It is a field at the intersection of machine learning and linguistics.
Its goal is to extract information and meaning from textual content.
\newline Among the text processing techniques that we exploit in our project, we cite \cite{mohan_preprocessing_2015} :
\begin{itemize}[label=\textbullet, font=\LARGE]

    \item \textbf{Empty word removal :} Empty word removal is one of the basic preprocessing
operations in various automatic language processing applications. The principle
consists in eliminating words that appear in the
commonly in all the documents of the corpus. 
\newline An empty word can be defined as a word that does not add meaning to the text and/or
its deletion does not change the meaning of the sentence.
\newline Generally speaking, articles and pronouns are usually classified as empty words.

\item \textbf{CamelCase : }This is a writing standard commonly used in programming languages.
Indeed in the names of variables or classes the words are not separated by spaces but
rather to indicate the beginning of the second word
we mark it with a capital letter.
\newline Therefore, we use this same principle to separate words in the document. For example
the word CamelCase will be : [Camel, Case].

\item \textbf{Stemming Vs Lemmatization : }In linguistics, stemming or desuffixation is a
process of reducing words to their root. 
\newline The root of the word is the part left after we
proceed by deleting
its prefix(es) and suffix(es).
\newline Unlike the lemma, which corresponds to a term taken
from the ordinary usage of the speakers of the language, the root generally
corresponds only to a term derived from this kind of analysis as illustrated by the
attached figure 2.2. 

\begin{figure}[H]
    \centering
    \includegraphics  [bb=0 0 15cm 8cm]{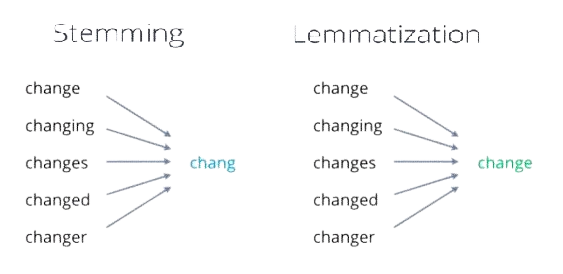}
    \caption{Stemming Vs Lemmatization}
    \label{Fig:Stemming Vs Lemmatization}
\end{figure}

\end{itemize}
\subsection{Term Frequency Inverse Document Frequency }
\paragraph{}
TFIDF is a technique used to vectorize data. It does this by proportionally increasing
the number of times a word appears in the document, but it is counterbalanced by the number of documents in which it is present. Therefore, words
that are commonly present in all documents, do not have a very high rank. However, a
word that is present too often in a small number of documents will be assigned a higher
rank because it may be indicative of the context of the document.\cite{liu_research_2018}
\newline We use this technique to vectorize our data and we'll detail it in the following chapters.

\subsection{Clustering}
\paragraph{} 
Clustering is essentially a type of unsupervised machine learning method, based
only on input data that is unlabeled. In other words, clustering is a process for finding
meaningful structure, generative features and inherent groupings in a set of examples.
\newline Clustering algorithms partition the data into a number of clusters (groups or categories). A
cluster is described by considering internal homogeneity and external separation. That is examples in the same cluster should be similar to each other (thus increasing cohesion
within the cluster), while those in different clusters should not (thus reducing the range
between clusters) \cite{xu_survey_2005}.
\paragraph{}
Density-based clustering is a non-parametric approach where clustering methods do
not require the number of clusters as input parameters. In this type of clustering, a cluster
is a set of example data distributed in the data space over a contiguous region of high
example density.
\newline However, density-based clusters are separated from each other by contiguous regions of
low object density. Thus, data objects located in the low density regions are generally
considered noise or outliers \cite{campello_density-based_2020}.
\paragraph{}
As mentioned, we consider this refactoring problem as a
clustering problem in which we have opted for different clustering algorithms especially
density-based ones.

\subsection{The graphs}
\paragraph{}
A graph is a non-linear data structure used to represent real world problems. A graph
consists of a finite set of nodes (vertices or points) connected to each other by edges.
These nodes can be in an arbitrary form forming an undirected graph or in the form of a
set of oriented pairs for a directed graph.

This type of data structure emphasizes the relationship between objects as well as between
the nodes of the graph. Therefore, the edges of the graph may contain a weight reflecting
the importance of the relationship or link between these two nodes. \cite{fortunato_community_2010}
\newline A main application of graphs is the detection of communities that we define in the
following lines.
\paragraph{}
A community can be defined as a subset of nodes that are loosely connected to each
other and loosely connected to nodes in other communities in the same graph. Detecting
communities in a network is one of the most important tasks in network analysis. In a
large scale network, we will need community detection algorithms that can partition the
network into multiple communities\citeURL{URL8}.
\newline There are mainly two types of methods for detecting communities in graphs :
\begin{enumerate}
    \item \textbf{Agglomerative methods: } In this approach, we start with an empty graph
composed only of the nodes of the original graph. Then, edges are added one by one
to the graph, starting with the edges of the highest weight (depending on the
application) \citeURL{URL8}.
    
    \item \textbf{Splitting methods :} In this method, we proceed in the opposite direction. We start with
the complete graph and remove the edges iteratively. The edge with the lowest weight is
removed first. After a number of steps, we obtain groups of densely connected nodes \citeURL{URL8}.
\end{enumerate}
In this project, we use graphs to represent our problem. Then, we apply community
detection algorithms in order to detect microservices. 

The introduction of these definitions of key concepts, which we use in the development
of our project, allows us to move on to the second part of the chapter through the first
research question. \vspace{0.5cm}

\textbf{ \large Research Question 1 :} What refactoring approaches exist in the context of
decomposing a monolithic application architecture into its microservices and how can they be classified based on the techniques used?

\section{Related work}
\paragraph{}
After defining the key concepts and to answer our first research question presented above,
in this section we start by explaining the research methodology we have used, then we present
the different existing solutions classified according to the approach used and we end with a
discussion.

\subsection{Research Methodology}
\paragraph{}
During our research process we followed the PRISMA approach which we present in
the next section because it is simple to apply and successful in conducting literature reviews.

\subsubsection{Presentation of PRISMA}
\paragraph{}
PRISMA is an abbreviation for Preferred Reporting Items for Systematic Reviews and
Meta Analysis. The process consists of a set of evidence based reporting items for
systematic reviews and meta-analyses. PRISMA is primarily designed to report on reviews
that assess the effects of interventions, although it can also be used to describe systematic
reviews with objectives other than the evaluation of interventions.\cite{page_prisma_2021}

\subsubsection{Application of the PRISMA method}
As shown in Figure 2.3 the steps in the PRISMA process are :
\begin{itemize}[label=\textbullet, font=\LARGE ]
    \item \textbf{ Identification phase :} This is the first step in the PRISMA process and it reintroduces the step of collecting related work based on keywords that match the
work. \newline We searched for research articles using a combination of a set of keywords on
Google scholar \citeURL{URL7} as a search engine, since it allows users to access the entire
scientific literature from a single location. \newline Connected papers \citeURL{URL14} was also a useful tool as it helped us to select the articles
connected to our research theme by inserting a first article representing our research
interest and generating all the articles semantically connected to it.
\newline The date range was from 2014 to 2022 using the key phrases: "Monolith
refactoring", "Monolith migration", Monolith to microservices", "GNN for Monolith decomposition", "Machine Learning for monolith refactoring", "Microservices based application", "Graph community detection".
\paragraph{}    
   For this step, we encountered 44 items. Five of them were deleted because they were
duplicate items.
    \newline In this first phase, despite the limited time frame of the project, we have tried to
select the most relevant research articles from among the most recent ones. Even if
we can assume that this work is not exhaustive, the selected research works give a
clear overview of the approaches already developed to solve this problem.
    
    \item \textbf{Selection phase :} This is the filtering stage of the selected articles. The first
selection criterion is based on the title and the abstract, depending on whether or not
they correspond to the project. The second criterion is the availability of the full text
of the article.
    \newline In the end, a complete reading of the articles is done to evaluate their eligibility. 
    \newline In
our situation :

    \begin{itemize}[label=\textbullet, font=\LARGE ]
        \item  8  items were excluded based on the Title/Abstract sort.
        \item   6  were not recovered because we were unable to find their full texts.
        \item 5 reports were excluded after review of their full texts because the approaches
used were not clearly explained.

    \end{itemize}
    
    \item \textbf{Inclusion Phase : } This is the reading and vetting stage of the articles that have met
all of the previously mentioned screening criteria.
\newline In our case, 20 articles were included in the review. On the one hand, 11 of them
contain generalities on the key concepts and the field of application. On the other
hand, the other 9 articles present recent approaches dealing with similar issues. The
latter nine research articles are used in the following comparative study.

\end{itemize}

\begin{figure}
    \centering
    \includegraphics [bb=0 0 15cm 8cm]{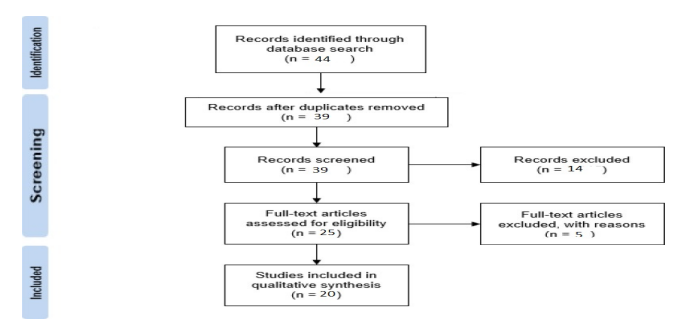}
    \caption{PRISMA-inspired flowchart for article retrieva} \protect \cite{hall_prisma_2020}
    \label{fig:PRISMA-inspired flowchart for article retrieva} 
\end{figure}

\subsection{Existing approaches}

In this section, we present the work done using the different articles selected through the
PRISMA filters.
\newline For that, we divide them according to the types of analysis that they carry out during the
realization of their solutions. Finally, we examine the extent to which these methods were
able to solve the project problem and conclude with our contributions. 

\subsubsection{UML diagrams}

The Unified Modeling Language (UML) is a graphical design language based on
diagrams and is designed as a standardized modeling method for software development
and object oriented design \cite{koc_uml_2021}.
\paragraph{}
Service Cutter is a very special migration tool as it is the only one in the literature that
is based only on UML diagrams to represent the different aspects to be treated of the
monolithic application to be decomposed. Starting from these diagrams, which are
rewritten in json format and provided by the user, Service Cutter generates a weighted
undirected graph representative of the application whose nodes are nanontities. The latter
are operations or methods and the edges are the interactions that take place. This graph is
generated based on 16 coupling criteria that Gysel and al.\cite{aiello_service_2016}  stated distilled from industry
experience. This decomposition tool also allows the expert user to order his criteria
according to his preferences. In a final stage of the microservices recommendation
process, Gysel et al. apply Girvan Newman's community detection algorithm and finally
generate a decomposition recommendation by defining the services and the nano-entities that
belong to it. It should be noted that this approach works on the basis of 9 software
diagrams that touch several aspects. Its 9 diagrams are : Entity-relationship diagram, Use
cases diagram, Shared owner groups, Aggregates, Entities, Predefined services, Separated
security zones, Security access groups, Compatibilities.

The problem with this approach is that most of the applications that exist are missing
representative diagrams and therefore it is up to the user to use reverse engineering
techniques to generate them and then transform them into the proper format accepted by
this approach. In addition, in terms of comparison of results, the developers of the Service
Cutter tool do not use known metrics in migration literature but rather rely on comparison
with a manual decomposition process that relies on domain experts.
\paragraph{}
Nevertheless, Dehghan and al.\cite{dehghani_facilitating_2022} have based their recently published approach on the
Service Cutter tool described above. Indeed, the user provides the necessary models and
the source code to two independent mechanisms:
\newline On the one hand, the Service Cutter tool receives the different diagrams in order to
generate a recommendation of the microservices-nanoentities.
\newline On the other hand, MoDisco (Model Driven Reverse Engineering Framework) \cite{bruneliere_modisco_2014}  receives
in turn the Entity-Relational model and the source code to generate a recommendation of
the corresponding methods-nanoentities. This approach is based on reinforcement learning
where the goal is to associate each method to its microservice according to the pre-results
previously synthesized by the two separate tools.

Like the Service Cutter, the verification of the quality of the results provided by this
approach was based on a manual process since the applications they used are small. It
should also be noted that for large systems, the system takes considerable time to stabilize.

\subsubsection{Static analysis of the code }

In computer science, the notion of static program analysis encompasses a range of
techniques for collecting information about the behavior of a software program during its
execution, without having to execute it.
\paragraph{}
In this quest  Brito and al.\cite{brito_identification_2021} have developed a somewhat particular approach where the
source code is considered as a text to be processed from which they have extracted
information in two forms:
\newline The first part is lexical by detecting all significant words in the source code and structuring
them with an abstract syntax tree (ASA) and the second part is structural by using
JavaParser Symbol Solver to identify the type of each given expression.
The lexical data
are used as input to the topic modeling mechanism which combines its output with that of the structural analysis to feed the clustering phase
that generates the final clusters. The results of this approach were found to be acceptable.
\paragraph{}
In the other end of the approaches based on static analysis of the code we present the
solution entitled "HierDecomp". We were inspired by the article of this last approach to develop our third
research track that we detail in the following chapters.\\
This article published by Sellami and al.\cite{sellami_hierarchical_2022} is based on the treatment
of the problem from a clustering point of view. According to their approach on
two kinds of metrics that they defined as the structural similarity synthesized from the
static calls between the classes of the application and the semantic similarity derived from
the analysis of the code text (we detail these two metrics in the third chapter of our report
when presenting our third approach). Candidate microservices are generated following a
hierarchical clustering done on the basis of these last metrics.
Sellami et al have shown
good results when compared to several distilled approaches in the literature.

\subsubsection{Dynamic code analysis}
\paragraph{}
Dynamic computer product analysis is a form of program analysis that requires the
execution of programs. It is intended to study the behavior of software and the results of
its execution on its environment.\cite{da_costa_exploring_2021}\\
Monolithic application migration approaches that rely on dynamic source code analysis
have provided better results.

 Among which we mention the proposal of the IBM research team in 2021, Desai and al \cite{desai_graph_2021}, who proposed an approach in which they take advantage of the advancement of
research in terms of GNN techniques. It is a method that treats the problem as a class-level
clustring

They start by generating an unweighted directed graph representative of the application
whose nodes represent the classes and edges represent a relationship between them. The
vectors that encode the nodes of the graph are synthesized thanks to the information
extracted from the application's execution traces.

The biggest advantage of this approach is that at the end of the process the system not
only generates candidate microservices but also shows the user which classes in each
cluster have a high interaction rate with the rest of the classes. According to the literature,
this type of class requires a lot of attention during refactoring. The results of this GNN based approach outperform the rest of the graph-based methods but to take advantage of
this solution the user has to provide the number of clusters as input which is not a simple
task.
\paragraph{}
However, and in the same context, IBM has recently published a refactoring tool called
"Mono2Micro", which is known to be the most efficient in the industry. Kalia and al. \cite{kalia_mono2micro_2021} have defined 6 steps for this migration process. They start with the collection of execution
traces due to the source code provided by the user. Then, for each use case, they
proceed to the reduction of these traces: First, by considering only unique traces. Second,
by reducing the length of a trace by removing redundant sequences of classes that could
have been invoked due to the presence of a loop.
\newline After a series of processing, the team ends up applying a hierarchical clustering to extract
the microservices and the user has access to the different levels of clustering.
\newline IBM also offers an interactive interface through which the user can visualize the final
result and can associate a class with another microservice of his choice different from the
one assigned to him by the decomposition tool and continue to compare the results he
obtains.
\newline Mono2Micro is the excellent decomposition tool in the monolithic application
migration industry.

\subsubsection{Decomposition patterns }
Sam Newman being an expert in the field at the industrial research level defines in his
book entitled “ "Monolith to Microservices Evolutionary Patterns to Transform Your
Monolith “ the different patterns that help in the decomposition of monolithic applications\cite{newman_monolith_nodate}.
\newline The most famous of these patterns is called the "Strangler Pattern" or "The Fig Principle".
This strangler pattern was inspired by a certain type of fig that is planted on the upper
branches of trees. The fig then descends to the ground to take root, gradually enveloping
the original tree. The existing tree initially becomes a support structure for the new fig. In
the final stage, we can observe the original tree die and rot, leaving only the new, now self sustaining fig in its place.
\paragraph{}
In the context of software, the parallel principle is that our new microservices-based system
is initially supported by the existing system and the wrapper.
\newline  The idea is to have the old and the new coexist, giving the new system time to develop and
fully replace the old system. The key advantage of this model is that it supports the goal of
enabling an incremental migration to a new microservices-based system. In addition, it
provides the ability to pause and even stop the migration completely, while continuing to take
advantage of the new system delivered so far.
\paragraph{}
As shown in figure 2.4, the process is simple:
We start by designating the functionality
to be extracted (to copy the code if possible) from the running monolithic application,
which corresponds to the first step in the figure. During the extraction of the functionality,
the calls are always directed to the parent monolithic application to ensure that the services
are not broken during this phase, this is illustrated by the second step of the process. After
the extraction phase is over and the functionality has become a full-fledged functional
service, the calls concerning it are then directed to the new microservice that represents it
and so on. This is represented through the last step in our figure 2.4.
\newline The end of our process is marked by the degeneration of the parent monolithic application.

\begin{figure}
    \centering
    \includegraphics [bb=0 0 15cm 6cm]{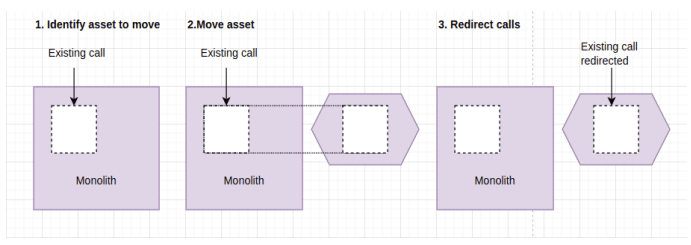}
    \caption{Illustration of the strangler pattern} \protect \cite{newman_monolith_nodate}
    \label{fig:Illustration of the strangler pattern} 
\end{figure}

\subsubsection{Discussion}
There has been considerable interest in service-oriented or microservice architectures
that allow for rapid system modification. Refactoring and, in particular, remodularization
operations can be performed to repair the design of a software system. Various approaches
have been proposed to help developers when remodularizing a software system. The
common problem in these efforts is to identify from monolithic applications the candidates
for microservices, i.e., programs that can be transformed into cohesive, self-contained
services. There are various decomposition approaches that in this section we have tried to
gather according to the type of analysis done by presenting their general aspects. 
\newline In order to take advantage of these approaches to build our own contribution, we need
to define an evaluation tool and this is what allowed us to pose our second research
question.
\paragraph{}
\textbf{ \large Research Question  2 :} In view of the different approaches extracted from the
literature that are the subject of solutions for our problem, How can we
evaluate these solutions and decide on their robustness ?

\section{Evaluation metrics}
\paragraph{}
In this section and in order to answer the second research question we have raised, we
present the evaluation metrics we use in the evaluation process of approaches including
ours so that we can position ourselves in this research quest. These metrics are distilled
from the literature mainly from the research work of IBM, the industry and research leader
in the field of monolithic application migration through its famous paper Mono2Micro \cite{kalia_mono2micro_2021} dating back to 2021 and whose approach we detailed in section 2.2.2 during the
presentation of existing approaches.
\newline The metrics we consider are:
\begin{itemize}[label=\textbullet, font=\LARGE]
    \item {\textbf{Structural Modularity (SM) : } Structural modularity measures the quality of
partition modularity as the structural cohesion of classes within a partition $m_i$ (scoh)
and the coupling (scop) between partitions(M), represented
by equation 2.1: 
    \begin{equation}
    { \frac{1}{M} } { \sum_{i=1}^{M} scoh_i} - { \frac{1}{(M(M-1))/2}} {  scop_{ij} }
    \end{equation}
    
    With :
    \begin{itemize}
        \item \emph{$ scoh_i $ }= $ \frac{\mu_i}{m_i^2} $,

        \item \emph{ $\mu_i$} represents the number of calls within the partition $m_i$,
        
        \item \emph{ $ scop_{ij} $ } =  $\frac{(\gamma_{ij})}{2*(m_i*m_j)}  $,
        
        \item \emph{ $ \gamma_{ij}$}represents the number of calls between partitions $m_i$ and $m_j$. 
        
    \end{itemize}

    \textbf{ →  The higher the SM value is, the better the recommendation..}
 
    }

    \item {\textbf{ICP:}  Measures the percentage of calls occurring between two partitions, represented
by Equation 2.2 : 
    \begin{equation}
    icp_{ij} = \frac{c_{ij}}{ \sum_{i,j=0 j/=i}^{M} c_{ij} } 
    \end{equation}
    
    with :
    
    \begin{itemize}
        \item  \emph{ $c_{ij}$} represents the number of calls detected between partition i and j.
    \end{itemize}

     \textbf{ → The lower the ICP value is, the better the recommendation.}
    } 
    
    \item {\textbf{Interface Number (IFN) : } Measures the number o f interfaces in a microservice. An
interface is defined as a class of a microservice $m_i$ that is called by another class of
another microservice $m_j$.
     This metric can be calculated
using equation 2.3 :
    \begin{equation}
        {\frac{1}{N}} { \sum_{i=1}^{N} ifn_i }   
    \end{equation}

     With :
     \begin{itemize}
         \item \emph{N}  being the total number of microservices, 
         \item \emph{$ifn_i $} represents the number of interface classes in the microservice $m_i$.
     \end{itemize}
     
     \textbf{ → The lower the IFN value is, the better the recommendation.}
    }
    \item {\textbf{Non-Extreme Distribution (NED) : } This metric reflects the way classes are
distributed within microservices. It is preferable that a microservice be neither very
large nor very small in terms of the number of classes it contains.
    Wu and al.\cite{wu_comparison_2005} define a microservice as non-extreme if its size varies between [ 5 , 20 ] classes. This value is measured by equation 2.4 :
    
    \begin{equation}
        1 - {\frac{\sum_{k=0}^{N} n_k}{|N|}}
    \end{equation}

    With :
    \begin{itemize}
        \item \emph{$n_k$}  represents the microservice classified as non-extreme,
        
        \item \emph{N} being the total number of microservices. 
    \end{itemize}
    
     \textbf{ → The lower the NED value, the better the recommendation.}
    }
    \item {\textbf{DUP : }This is another metric that controls the size of microservices by calculating the
number of duplicated classes. Based on the fact that one of the solutions to reduce the
coupling between partitions is the duplication of the co-used classes.
\paragraph{}
     \textbf{ → The lower the DUP value is, the better the recommendation.}
    }

\end{itemize}

\textbf{\Large Conclusion}
\paragraph{}
In this chapter, we started by enumerating the key concepts necessary for a good
understanding of our project. Then, in a second part, we presented the PRISMA research
methodology that we followed when selecting the scientific articles that we exploited in
the existing approaches part, followed by a discussion in order to answer the first research
question elaborated previously.
\newline Finally, and to answer the second research question, we have closed this chapter by
detailing the evaluation metrics used to compare and position the approaches, including
our own, which we will present in the following chapters.

%% file: Chapitres/chapitre3.tex
\chapter{Proposed approaches}
\label{chap:chap3}

\textbf{\large Introduction}
\paragraph{}
In this third chapter of the report, we start by presenting the data understanding phase
in a first section while mentioning the tools used to satisfy this task. Then, in a second
section, we detail the proposed approaches by focusing on the data preparation part, the
second phase of the CRISP-DM process.

\section{Data understanding}
\paragraph{}
We treat this refactoring problem as a clustering problem. Our solution takes as input
the source code of the applications to be split into microservices. The level of granularity
we consider is at the class level.
\paragraph{}
\textbf{A class :} A class in the Java object-oriented programming language is a basic element
that can be introduced as a definition model for objects with the same set of attributes, and
the same set of operations. A class allows to create one or more objects by instantiating.
Each object is an instance of a single class \citeURL{URL9}.
\paragraph{}
Our solution does not replace the decision of the expert but recommends a way to
gather the classes of the application while presenting the different results of the evaluation
metrics that are suitable to evaluate the approaches to solve this migration problem and
that we presented in the previous chapter in section 2.3.

To do so, we collected two monolithic applications to train our data and test our
clustering approach.

Indeed, these applications have been chosen according to the following criteria :
\begin{itemize} [label=\textbullet, font=\LARGE ]
    \item Monolithic application written in Java.

    \item Availability of source code.
    
    \item To be used by other approaches so that we can compare approaches performances to the results of
different reference approaches.
    
\end{itemize}
However, the monolithic applications that form our database are:
\begin{itemize}[label=\textbullet, font=\LARGE]
    \item \textbf{ Daytrader \citeURL{URL10}: } This sample contains the DayTrader 7 benchmark, which is
an application built around the concept of a stock trading system
online. The application allows customers to log in, view their portfolios, check stock
prices and buy or sell shares. 
    
    \item \textbf{ Acmeair \citeURL{URL11} : }This application presents the implementation of a fictitious
airline called "Acme Air". The application was built with some key business
requirements in mind: the ability to scale to billions of calls of Web APIs per day, the need to develop and deploy the application in public
clouds (as opposed to a dedicated pre-assigned infrastructure), and the need to
support multiple user interaction channels. 
    
\end{itemize}
We summarize in this table the characteristics of the applications mentioned above:
\begin{table}[h]
\centering
\begin{tabular}{|r|m{3cm}|m{3cm}|m{3cm}|m{3cm}|}
\hline  Project & Version & Number of lines
in the
code & Nbr of classes \\

\hline    Acmeair \citeURL{URL11} & 1.2 & 8,899 & 86 \\

\hline    DayTrader \citeURL{URL12} &  1.4 & 18,224 & 118 \\

\hline
\end{tabular}
 \caption{ Characteristics of monolithic applications}
  \label{- Characteristics of monolithic applications}
\end{table}

After selecting the monolithic applications and retrieving their source codes, we start
the data collection process. Indeed, our approach is based on the calls between the classes
of the application. We perform a static analysis of the source code. To do this, we use a
static code analysis software called "Understand".
\paragraph{}
 \textbf{\large Understand} 
 \paragraph{}

Understand, one of SciTools' products, is a customizable integrated development
environment (IDE) that enables static code analysis through a set of visual, documentation
and measurement tools. It is designed to make it easy for software developers to
understand, maintain and document source code. It enables code understanding by
providing flowcharts of relationships and building a dictionary of variables and procedures
from a provided source code \citeURL{URL15}.
\paragraph{}
In our project, during the data collection phase, we use the API provided with
Understand to generate the class dependency matrices for each of the applications by
considering only the calls between classes.
\newline We detail the generated dependency matrices in the next part of our chapter.
%\vspace{-2cm}
 
\section{Research avenues explored}

During the development of our problem solving approach, we conducted three rounds
of the CRISP-DM work methodology mentioned in the first chapter of our report. As a result, we explored three avenues of research discussed in this section.

\subsection{First approach : Naive method }
In this first approach, we start by generating the dependency matrix between the
classes of the "Acmeair" application as shown in figure 3.1 :

\begin{figure}[H]
    \centering
    \includegraphics [bb=0 0 12cm 6cm]{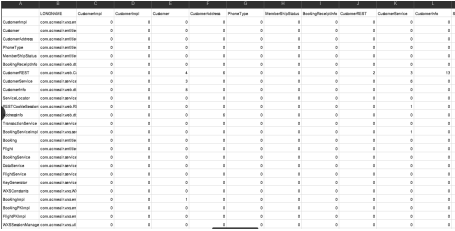}
    \caption{ Extract of the dependency matrix between the classes of the Acmeair
application }
    \label{Fig: Extract of the dependency matrix between the classes of the Acmeair
application }
\end{figure}
From this matrix, we generated a tuple to represent each class of the application in
question in the form [ a , b ]  such that : 
 \begin{equation}
      a = \sum call_{in}  
 \end{equation}
 
 \begin{equation}
     b = \sum call_{out} 
 \end{equation}
 
 With :
 \begin{itemize}
     \item \emph{$call_{in} $} represents the incoming calls from all other classes to the class in question, 
     
     \item\emph{ $ call_{out}$} represents outgoing calls from the class in question to all other classes in the project.
 
 \end{itemize}
 
 In the end, the data we process in this first approach are only a representation of each
class of the application formed using equations (3.1) and (3.2).
 \paragraph{}
 The idea behind this choice of representation is indeed due to the observation that in
the dependency matrices, we find classes that interact a lot in terms of outgoing calls but
that these same classes are also called several times by other classes. In other words, after
this observation, we want the classes that interact a lot to be in the same cluster in order to
reduce the coupling between the clusters and to increase the cohesion.

\subsection{Second approach: Codependent call method}
\paragraph{}
In this second approach, we start by generating the matrix of dependencies between the
classes illustrated in the previous figure 3.1 but this time we refine the metric considered
when generating the class representations.\\
Indeed, in this approach we consider codependent calls between classes.
To make this notion clear, we proceed by the example of the attached figure 3.2 : 

\begin{figure}[H]
    \centering
    \includegraphics [bb=0 0 15cm 8cm]{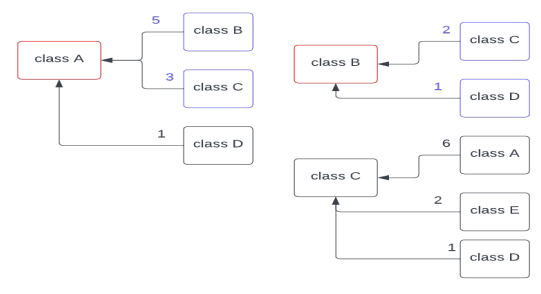}
    \caption{Illustrative example of the metric considered  }
    \label{Fig:Illustrative example of the metric considered }
\end{figure}

Considering five classes A, B, C, D, and E. Our goal in this example is to encode class
A.

The figure shows that class A is called 5 times by class B, 3 times by class C and a
single time by class D.

We can notice that the classes that commonly call classes A and B are classes C and D. 
Also the classes A and C have in common a single class D. 
Therefore, class A is encoded according to class B (i.e. according to the class that has a
maximum of incoming calls from the same classes as those calling class A).
\paragraph{}

In other words, we consider a tuple representing the class A in the form [ a , b ] such that  :
 
\begin{equation}
    a =  \sum call_i  \\
\end{equation}

\begin{equation}
b=  \frac{call_{in} A}{call_{in} B}  \\
\end{equation}
With :
\begin{itemize}

    \item \emph{$call_i$}  : : represents the calls of codependent classes between A and B,
    
    \item \emph{$call_{in} A$ }:  is the set of all incoming calls to A.
    
    \item \emph{$call_{in} B$}: is the set of all incoming calls to B.
\end{itemize}

Finally, the line that will represent the class A is none other than the combination
of the two equations (3.3) and (3.4). 
 \paragraph{}
 This representation metric was inspired by the article "Mining Multi-level API
Usage Patterns"  \cite{saied_mining_2015} dealing with the theme of API usage pattern extraction. We are
inspired by this same metric used for our second research track for our monolithic
application migration problem.
 \paragraph{}
 The idea behind this choice of representation is to gather the classes that are called
together in the same microservice and this will guarantee not only a weak coupling
between the resulting microservices but also the criterion of only one use case invoked by
microservice.

\subsection{Third approach : Graph method}
\paragraph{}
During this last approach, we based on the research paper entitled "A Hierarchical DBSCAN Method for Extracting Microservices from Monolithic Applications" \cite{sellami_hierarchical_2022} dealing with the theme of our project\\
In this approach, we no longer encode each class separately but rather the relationship or
link between pairs of classes.
\paragraph{}
For the purpose of extracting relationships and dependencies between different classes,
we define two distinct types of similarity between classes as follows :

\begin{itemize} [label=\textbullet, font=\LARGE]
    \item \textbf{Structural similarity  ($sim_{str}$)} : This metric is based on the shared number of
method calls between 2 classes. It encodes the dependency between them and thus
evaluates the similarity from a functional point of view. The objective of the grouping
of classes based on this similarity is to obtain consistent microservices from an
implementation point of view.
\paragraph{}
Considering two given classes  $ C_i $ and $ C_j $,   the structural similarity is defined by equation (3.5) :

\end{itemize}
    
\begin{equation}
sim_{str}(c_i,c_j)=  
	\left\lbrace  
		\begin{aligned}
			\frac{1}{2}{(\frac{call(c_i,c_j)}{call_{in}(c_j)} + \frac{call(c_i,c_j)}{call_{in}(c_i)} )} \hspace{0.25cm} if \hspace{0.25cm} call_{in}(c_i)  \ne 0  \hspace{0.25cm} and \hspace{0.25cm} call_{in}(c_j) \ne 0\\
            \frac{call(c_i,c_j)}{call_{in}(c_j)} \hspace{0.25cm} if \hspace{0.25cm} call_{in}(c_i) = 0 \hspace{0.25cm} and \hspace{0.25cm} call_{in}(c_j) \ne 0\\
            \frac{call(c_i,c_j)}{call_{in}(c_i)} \hspace{0.25cm} if \hspace{0.25cm}call_{in}(c_i) \ne 0 \hspace{0.25cm} and \hspace{0.25cm} call_{in}(c_j) = 0	
		\end{aligned}
	\right\}
\end{equation}

With:
\begin{itemize}
    \item call( $c_i$,$c_j$ ) :   represents the number of times a method of class $c_i$ i has called a
method of class  $c_j$.
    
    \item $call_{in} $($c_i$) :  represents the number of incoming calls in $c_i$.
    
\end{itemize}

The values of $ sim_{str}$ ($c_i$,$c_j$ )  are in the interval [0,1] où 1 where 1 indicates that the
classes $c_i$ and $c_j$ j are very similar and used exclusively together and 0 indicates
that they are completely independent.

To calculate this metric, we used the dependency matrices generated by
Understand.
\begin{itemize} [label=\textbullet, font=\LARGE ]
    \item \textbf{Semantic Similarity  ($Sim_{sem}$) : }This metric uses natural language processing
(NLP) to measure the degree of relatedness of the domain semantics of two given
classes. Given that a microservice must provide a specific and/or a single use case in the domain, we need to identify classes that serve
similar use cases.
\newline Assuming that monolithic projects have been coded using standard practices where
class, method and variable names reflect functional concepts and comments
describe their function. The terminology used within these components can be a
powerful tool for extracting the domain meaning of each class.\\
Thus, each class is defined by the set of words contained in its comments, parameter
names, method names and variable names as illustrated in figure 3.3 where the first
column represents the name of the class and the last column encloses the list of
extracted words : 

\begin{figure}[H]
    \centering
    \includegraphics [bb=0 0 15cm 3cm]{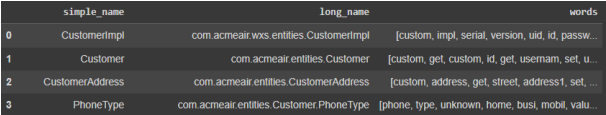}
    \caption{ Extract from the semantic analysis of the Acmeair project}
    \label{Fig:- Extract from the semantic analysis of the Acmeair project }
\end{figure}

Each word is pre-processed using natural language processing techniques
including splitting using CamelCase, filtering out unnecessary words and
stemming.\\
The final result is a vector of size  $n_v$ where $n_v$ is the number of words in the
domain extracted from the monolithic application. Finally, this vector is measured
using a TF-IDF (Term Frequency-Inverse Document Frequency) model.\\ 
The semantic similarity metric is defined as the cosine similarity between two
classes, as presented by equation (3.9) :

\begin{equation}
    Sim_{sem}(c_i;c_j)=\frac{\vec{c_i}.\vec{c_j}}{||\vec{c_i}|| . ||\vec{c_j}||}
\end{equation}

With:
\begin{itemize}
    \item $\vec{c_i}$ and $\vec{c_j}$ :  represent the TF-IDF vectors of class $c_i$ and class $c_j$.
    \item $|| \vec{c_i}||$ : represents the Euclidean norm of the vector  $\vec{c_i}$.
\end{itemize}

The values of   $ Sim_{sem}$ are between 0 and 1, where the value 1 means that both
classes use exactly the same vocabulary and therefore serve the same use case.
\paragraph{}
\item\textbf{Class Similarity (CS) :} The previous similarity metrics represent different aspects
of the relationships between classes. These two metrics are not necessarily
correlated and using only one of them does not guarantee the
satisfaction of the other. For these reasons, we use the class similarity metric which
represents a weighted sum between the structural similarity and the semantic similarity of two given classes presented by equation (3.10) : 

 \begin{equation}
     CS(c_i,c_j)= \alpha Sim_{str}(c_i,c_j) + \beta Sim_{sem}(c_i,c_j)
 \end{equation}

With : 
\begin{itemize}
    \item \emph{$\alpha \in $ [0,1]}, 
    
    \item \emph{ $\beta \in $ [0,1]}, 
    
    \item \emph{$\alpha$ + $\beta$ = 1.}
\end{itemize}

\end{itemize}
\paragraph{}
\textbf{\large Conclusion}
\paragraph{}
In this chapter we have started by presenting the data understanding phase
corresponding to our project by defining the level of granularity considered when
solving the migration problem as well as shedding light on the monolithic applications that
we consider for the treatment of the problem.
\newline Then, after defining that our approach is classified under the section of solutions dealing
with the problem using static analysis of the code, we highlighted the static analysis
tool that we used during our research.
\newline Finally, and during the second part of this chapter, we swept through the three research
avenues we explored while highlighting the data preparation phase of the CRISP-DM
methodology.

%% file: Chapitres/chapitre4.tex
\chapter{Modelling
}
\label{chap:chap4}

\textbf{\large Introduction}
\paragraph{}
After having collected and pre-processed the data for each of the approaches explored
in the previous chapter, we focus in this part of the report on the modeling phase of the
CRISP-DM process. To do so, we detail in this fourth chapter the algorithms developed for
each of the three approaches. 
\section{Modeling the naive approach: }
\paragraph{}
On the basis of the data that we select and reserve for this first solution and as we treat
this problem from a clustering point of view we have thus applied two clustering
algorithms on this last input.\\
The two chosen algorithms are DBSCAN and BMSC. In the following section, we detail each of these two algorithms. 
 \subsection{Density-Based Spatial Clustering of Applications with Noise }
 \paragraph{}
The first algorithm used is Density-Based Spatial Clustering of Applications with Noise
(DBSCAN).
\subsubsection {Generalities and benefits}
\paragraph{}
DBSCAN is a density-based clustering algorithm used to cluster data of any shape in the
presence of outliers and noise in a large amount of data.\cite{singh_literature_2022} 
\newline As shown in figure 4.1, the main advantage of the DBSCAN algorithm over partitioning
and hierarchical clustering algorithms are:
\begin{enumerate}
    \item The DBSCAN algorithm can be applied to any arbitrary shape of data rather than
spherical or convex shapes. This makes it much more practical than k-means and
other clustering algorithms.
   
   \item It has more specific advantages over the k-means algorithm, such as not having to
specify the value of k which is the number of clusters, so the number of clusters
can be arbitrary depending on the data which makes the clusters larger and more
accurate.
   
   \item The major advantage of this algorithm, as its name suggests, is that it is also effective
in the presence of noise and outliers.
   
  Figure 4.1 shows the difference in clustering results between DBSCAN and Kmeans on different shaped datasets where each color in the figure represents a
cluster.
\end{enumerate}
\begin{figure}[H]
    \centering
    \includegraphics[bb=180 0 3cm 5cm] {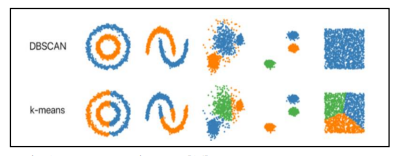}
    \caption{DBSCAN vs K-means  \cite{singh_literature_2022}}
    \label{Fig:DBSCAN vs K-means}
\end{figure}
\subsubsection{Hyperparameters}
\paragraph{}
DBSCAN is designed to discover clusters and noise in a spatial database. The
researchers need to know the parameters Eps and MinPts . After that, it gathers the
densely connected points together in a single cluster using the parameters already
entered.
\newline DBSCAN forms the cluster using 2 parameters Eps and MinPts:
\begin{itemize}[label=\textbullet, font=\LARGE ]
    \item \textbf{Eps ($\epsilon$) :}  This is the neighborhood radius around the point selected as the center of
the cluster to be formed. It is therefore used to determine the densest region. \\
 \textbf{→  If a large Eps value is chosen, DBSCAN will form less dense clusters and
thus the number of detected clusters will decrease.  } 
 
 \item \textbf{MinPts :}  This is the minimum number of points required to form a cluster.\\
 \textbf{→  The MinPts parameter should not be very low. }
\end{itemize}

\subsubsection{ Types of detected points}
\paragraph{}
After the completion of the DBSCAN algorithm on any dataset we get mainly 3 types of points as shown in figure 4.2 :
\begin{itemize}[label=\textbullet, font=\LARGE ]
    \item \textbf{Core point :} : This is the point from which there are at least a number "k" of points
(MinPts) at a distance "r" of radius (Eps).
    
    \item \textbf{Border point :}  This is any point that has one or more center points at a radius "r"
(Eps).

    \item \textbf{Noise :} Any point that is neither a core (center point or core) nor an edge point.\\
    In our project, we do not take into account the noise points detected by DBSCAN.
\end{itemize}

\begin{figure}[H]
    \centering
    \includegraphics[bb=180 0 3cm 5cm] {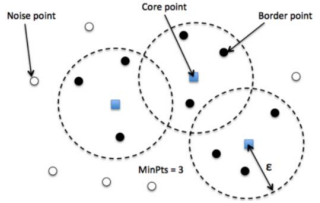}
    \caption {DBSCAN \cite{singh_literature_2022}}
    \label{Fig:DBSCAN }
\end{figure}
\subsubsection{Application steps }
Finally, the steps to realize the DBSCAN algorithm are as follows:
\begin{enumerate}
    \item Select an arbitrary point in the database and consider it as our first core point.
    
    \item The collection of data points that are at a maximum distance equal to Eps
    
    \item If the number of points collected is greater than or equal to the minimum number of
points required (MinPts), the cluster will be formed.
    
    \item To expand the cluster, these same steps will be repeated for each of the cluster points.
    
    \item At this stage the algorithm generates the first cluster. It will then delete all the
points that build it from the database and start the same process again. The
algorithm stops when no more clusters can be formed considering the parameters
already provided. The remaining points will be classified as noise.
\end{enumerate}
 \subsubsection{Limites of DBSCAN}
 Despite the good results of the DBSCAN algorithm, it has some limitations such as :
 \begin{itemize}[label=\textbullet, font=\LARGE ]
 \item It has been observed that its performance is poor on variable density databases.
Indeed, the Eps parameter always reduces the selection of very distant points and
thus points with variable density which does not satisfy the
Minpts and therefore DBSCAN creates wrong clusters.
 
     \item It is highly user dependent as it takes 2 input values, namely Eps and MinPts. In
general, the user does not have a deep knowledge of the data which leads to
erroneous clusters.
     
     \item It causes some computational overhead that is not always appropriate for large
datasets.

 \end{itemize}
To conclude, after having obtained the experimental results that we present in the fifth
chapter of the report, we develop a second algorithm based on a theoretical paper entitled "Boosted Mean Shift Clustering (BMSC)" known to be
more robust than DBSCAN. In the following section, we detail the BMSC algorithm.
\subsection{Boosted Mean Shift Clustering  } 
\paragraph{}
BMSC is a clustering algorithm that we implement on the basis of its theoretical paper.
This last one comes to merge two clustering algorithms which are Mean shift and
DBSCAN that we have just explained its principle.\\
Indeed, Mean Shift is a non-parametric clustering technique that does not require the
number of input clusters and can find clusters of arbitrary shapes. Although attractive, the
performance of the Mean shift algorithm is sensitive to the selection of its parameters.
\newline DBSCAN is an efficient density-based clustering algorithm, but it is also sensitive to its
parameters and usually merges overlapping clusters. In this section, we propose Boosted
Mean Shift Clustering (BMSC) to solve these problems.\cite{calders_boosted_2014}\\
Adding to the previously described advantages of DBSCAN, BMSC comes to
override the limitations of Mean Shift and DBSCAN, while retaining their nonparametric nature. We seek to capture the underlying group structure of the data by
selecting a subset of the data that provides the skeleton of the clusters.\\
In order to explain how BMSC works, we start by detailing the Mean Shift. 
\paragraph{}

\textbf{\large Mean Shift \cite{derpanis_mean_2005}:}\\
The Mean Shift algorithm is a non-parametric clustering technique that estimates the
number of clusters directly from the data, and is able to find irregularly shaped clusters.
\newline It assigns data points to groups iteratively by moving the points to the mode. The mode
is the point representing the center of density in the selected region. This mode point is
detected by performing an estimate of the kernel density of the mixture in question as
shown in the attached figure 4.3.
\newline Points that converge to the same mode are considered members of the same group.
\newline The key parameter of the Mean Shift is then the kernel bandwidth. Its value can affect the performance of the mean shift and is difficult to define.

\begin{figure}[H]
    \centering
    \includegraphics [bb=0 0 10cm 7cm]{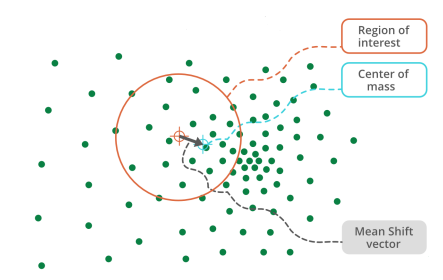}
    \caption{Mean Shift \protect \citeURL{URL12}}
    \label{Fig:Mean Shift }
\end{figure}

As shown in Figure 4.3, the steps for applying Mean Shift are in the order :
\begin{enumerate}
    \item  The selection of an arbitrary region (the region in red on the figure).
    
    \item The calculation of the center of density of the region (the blue point on the figure)
and the generation of the mean shift vector.
    
    \item  The displacement of the center of the region along the direction of the mean shift
vector until coincidence.
    
    \item Repeat steps 2 and 3 until convergence.

\end{enumerate}

\subsubsection { Hyperparameters of BMSC}
The BMSC hyperparameters are the combination of Mean Shift and DBSCAN
hyperparameters :

\begin{itemize} [label=\textbullet, font=\LARGE ]
    \item kernel bandwidth
    
    \item Eps 
    
    \item MinPts 
    
\end{itemize}

\subsubsection {Application steps}
The application steps of the BMSC algorithm can be illustrated through the
pseudo-code in figure 4.4:

\begin{figure}[H]
    \centering
    \includegraphics [bb=0 0 10cm 11cm] {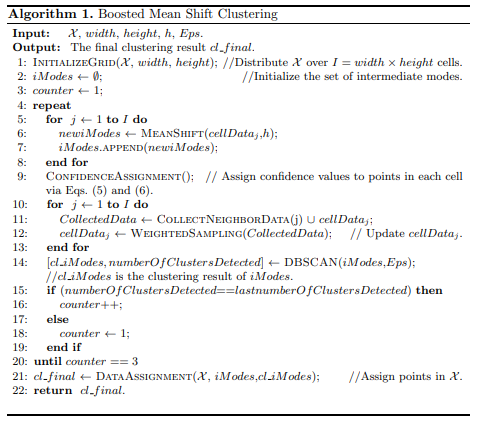}
    \caption{ BMSC algorithm \protect \cite{calders_boosted_2014}}
    \label{Fig:BMSC algorithm  }
\end{figure}

In this algorithm, we start by distributing the data evenly across the cells of a grid that
the user defines. In our project, we have defined a 3*3 grid.
\newline Then, in a first step we apply the Mean Shift algorithm on the data of each cell of the grid
and consequently we obtain a list of mode points called the intermediate modes
(iModes). after that, we redistribute the data of each cell according to the following
mechanism :
\paragraph{}
Each cell of the grid interacts with a finite number of cells in its surroundings,
which form its neighborhood according to a previously defined neighborhood structure.
The different neighborhood structures are illustrated in figure 4.5:
\begin{figure}[H]
    \centering
    \includegraphics [bb=0 0 10cm 3cm]{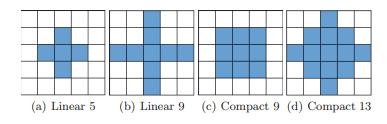}
    \caption{Neighborhood structures \protect \cite{calders_boosted_2014}}
    \label{Fig:Neighborhood structures  }
\end{figure}
In our project, given the size of our grid, we use the linear(5) neighborhood
structure.
\paragraph{}
Afterwards, in a second step of the re-sampling mechanism, we compute the distance
that separates all the points of the parent cell and its neighborhood from the iModes
detected by the Means shift. We assign the points to the cell depending on the distance
that separates them from the iModes of the cell while keeping the same initial cell size.\\
 At this stage, we assign each point to the iMode that is closest to it.
\newline Then we apply DBSCAN on the list of iModes in order to collect the iModes that are densely
packed and subsequently at a lower level we build clusters of the original data points. We
repeat these same steps until DBSCAN detects the same number of clusters for three
consecutive iterations.
\newline The benefits of BMSC are :

\begin{itemize}[label=\textbullet, font=\LARGE ]
    \item BMSC is recognized by better test results than DBSCAN. This will be proven in
the results that will be presented in the next last chapter.
    
    \item  The execution time is also acceptable compared to the computation time taken by
DBSCAN alone because the latter will only be applied on a reduced number of
data points and not on the whole database.
    
    \item The authors of BMSC proved in their paper that compared to Means Shift and
DBSCAN separately, BMSC is not as sensitive to the choice of hyper-parameters and
this by varying the hyper-parameters one by one and by controlling the evolution of
the results of the algorithm.

\end{itemize}

\section{Modeling the codependent call approach:}
\paragraph{}
In this second approach, we apply the two algorithms: DBSCAN in a first part and
BMSC in a second part.
\newline We will conduct a comparative study of the results in the next chapter of our report.

\section{Modeling of the graph theory approach}
\paragraph{}
This third and last approach models the problem using graph theory. Starting from the
weighted directed graph illustrated in figure 4.6 where the red nodes represent the
classes of the application and the edges represent a relationship according to the class
similarity metric detailed in the third chapter of the report.  
\newline To solve our problem, we apply two community detection algorithms that we present in the following
parts.
 
\begin{figure}[H]
    \centering
    \includegraphics [bb=180 0 5cm 9cm]{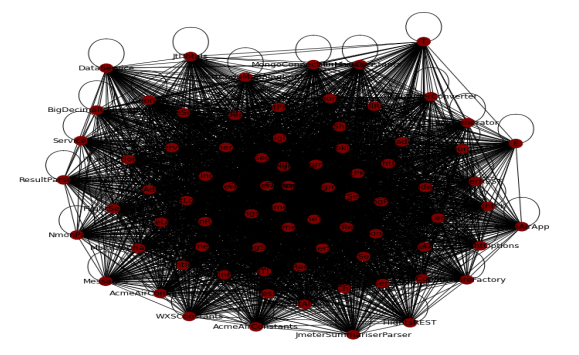}
    \caption{Weighted graph showing the relationships between the classes of the
Acmeair application }
    \label{Fig: Weighted graph showing the relationships between the classes of the Acmeair application  }
\end{figure}

\subsection{ Girvan-Newman }
The Girvan-Newman is an algorithm for detecting and analyzing community structure
by splitting relies on the iterative elimination of edges that have the largest number of
shortest paths between the nodes that cross them. By eliminating the edges of the graph
one by one, the network decomposes into smaller pieces, called communities.\cite{hurajova_revising_nodate} \\It can be summarized in the following steps :
\begin{itemize}[label=\textbullet, font=\LARGE ]
    \item For each edge of a graph, we compute the interdependence centrality of the edge.\\
Edge interdependence centrality is a measure of the centrality of an edge in a network
based on the number of shortest paths that pass through the given edge. It thus
identifies edges in the network that are crucial for in- formation flows.\\
In our case, these values are those of the class similarity that we have calculated
from the fact that classes that are similar to each other will have a higher class
similarity value is therefore an important flow of information that passes through the
edge that connects them.

\item We remove the edge with the smallest interdependence centrality.

\item We repeat the two previous steps until there are no more edges.
\end{itemize}

\subsection{Louvain}
\paragraph{}
Louvain is a community detection algorithm that is part of the agglomerative approaches.
It is a heuristic method based on the optimization of modularity. The algorithm works in
2 steps \cite{de_meo_generalized_2011} :
\begin{itemize}[label=\textbullet, font=\LARGE ]
    \item  Assign each node to be in its own community.
    
    \item Try to find the maximum positive modularity gain by moving each node to all
neighboring communities. If no positive gain is obtained, the node remains in its
original community.
\end{itemize}

\paragraph{}

\textbf{\large Conclusion}
\paragraph{}
In this chapter, we started by presenting the modeling phase which concerns each
of the developed resolution approaches by detailing the used algorithms.\\

%% file: Chapitres/chapitre5.tex
\chapter{Implementation and results}
\label{chap:chap5}

\textbf{\large Introduction}
\paragraph{}
In this last chapter, we present in a first section the realization environment of our
solution. Then, we show our results by comparing them to those distilled from different
approaches in the literature and we end with a discussion highlighting our contributions
during this project.

\section{Work environment}
\paragraph{}
In this section we present the hardware and software environment necessary to achieve
our goal.

\subsection{Software environment}
\paragraph{}
In this section, we present the different technologies used to implement this
project.
\begin{itemize}[label=\textbullet, font=\LARGE ]
    \item  \textbf{Python: } It is an object-oriented, interpreted, multi-paradigm and multi-platform
programming language. It is simple to master and can be used in a wide variety of
fields, including web development, data science,
AI, security, etc. Therefore, it is one of the most widely used languages\citeURL{URL16}. 
    
    \item \textbf{Scikit-learn (Sklearn) :}  This is the richest and most useful library for machine
learning in Python. It offers a range of powerful tools f o r machine learning and
statistical modeling, as well as classification, regression, clustering and dimensional reduction, through a
consistent interface in Python.\citeURL{URL17}. 
    
    \item \textbf{Natural Language ToolKit (NLTK) :} NLTK is a leading platform for creating
Python programs for working with data built in
from human language. It provides easy-to-use interfaces and a suite of text
processing libraries for classification, tokenization, stemming, tagging, parsing,
syntactic analysis and semantic reasoning. \citeURL{URL18}. 
    
     \item \textbf{Understand :}  It is a static analysis tool of the source code of applications allowing
to generate a set of information for our case we used it to generate the call matrices
between the classes of the application.\citeURL{URL15}.

\end{itemize}
\subsection{Hardware environment}
\paragraph{}
In order to implement our solution, we used different resources :
\begin{itemize}[label=\textbullet, font=\LARGE ]
    \item  \textbf{Google Colaboratory :  } It is a platform of free resources offered by Google. It has : 1 GPU (Tesla T4) With 15GB \citeURL{URL19}. 
    
    \item \textbf{ASUS : }For the development of the project, we used an ASUS personal computer with
these features :\\
Processor : Intel i7-7500U CPU 2.70GHz |  Operating System : Windows 10.
Mémory : 8 GB RAM | Hard Disk : 1 To.
\end{itemize}

\section{Experimentation}
\paragraph{}
In order to test the developed approaches and draw results from them, we must, at this
stage, fix the hyper-parameters of each approach.\\
We used the same hyper-parameters for the application of the first two approaches :

\begin{itemize}[label=\textbullet, font=\LARGE ]
    \item \textbf{ MinPts: } Let be the hyper-parameter corresponding to the DBSCAN algorithm. In
order to optimize our solution and to better satisfy the evaluation metrics, this
hyper-parameter has been fixed during all the experimentation to \textbf{MinPts=5}.\\
   Indeed, the NED evaluation criterion defined in Chapter 2 indicates that a micro
service is not considered very small if it contains at least five classes.
    
    \item \textbf{ Eps: } The limiting distance between two classes for one to be considered in the
neighborhood of the otheris the most important DBSCAN parameter to
choose appropriately for our data set.
\newline  We varied the Eps values from 0 to 1 with a step equal to 0.05 to obtain the best
results which we present in the next section for the value of \textbf{Eps = 0.65}.
    
    \item \textbf{ Kernel bandwidth : }This hyper-parameter was determined using the $estimated_bandwidth$ function which uses a heuristic based on the median of all
pairwise distances to generate the best one that matches our input data.
\end{itemize}

On the other hand, the hyper-parameters of the third and last approach have been
determined from the literature. Indeed, Sellami and al.\cite{sellami_hierarchical_2022} in their articles have
conducted a comparative study of the variation of the parameters in order to derive the
best value.\\
Therefore and based on their study, we set these hyper-parameters as follows :

\begin{itemize}[label=\textbullet, font=\LARGE ]
    \item \textbf{Alpha $\alpha$  : } Being the coefficient that represents the weight of the structural analysis
in the class similarity metric, hence in determining the relationship between each
pair of classes.\\
According to Sellami and al.\cite{sellami_hierarchical_2022} the optimal values for the hyper-parameter  $\alpha$ are in
the interval [0,45, 0,55].Therefore we have fixed \textbf{$\alpha$ = 0.5}.\\
This indicates that semantic similarity and structural similarity are equally
important in determining the valuation of the relationship between two classes.
    
    \item \textbf{Beta $\beta$ : } Therefore, by satisfying the condition $\alpha +\beta = 1 $ \textbf{$\beta$ }  we obtain \textbf{0.5}.

\end{itemize}
The algorithms that we have applied for the detection of communities within the graph
only require as parameters the graph and the list of weights of the edges.\\
After fixing the hyper-parameters of each of the developed approaches, we move to our
third and final research question :

\paragraph{}
\textbf{ \large Research Question 3 : }How does our solution compare to the different
approaches extracted from the literature that are being solved for our microservices
extraction problem?

\section{Results}
\paragraph{}
In this last section, we present the results of our solution, which corresponds to the
fifth phase of the CRISP-DM methodology entitled "Evaluation". We conclude this
section with a discussion comparing our final results with those of synthesized approaches
in the literature.
\paragraph{}
To begin with, we summarize in the two attached tables 5.1 and 5.2 the different
results of the state-of-the-art approaches applied respectively to the Acmeair and
Daytrader applications.
\newline The first column of the table contains the evaluation metrics that we presented in the
second chapter and the rest of the columns present the results of different approaches in
the literature.

%tableau acmeair littérature
\begin{table}[htbp]
\centering
\begin{tabular}{|r|m{2cm}|m{2cm}|m{2cm}|m{2cm}|m{2cm}|m{2cm}|}
\hline  Acmeair & Mono2Micro-IBM %\cite{kalia_mono2micro_2021}
& Bunch-HierDecomp %\cite{sellami_hierarchical_2022}
& GNN  %\cite{sellami_hierarchical_2022}
& FoSCI-HierDecomp %\cite{sellami_hierarchical_2022}
& MEM-HierDecomp %\cite{sellami_hierarchical_2022}
%& GNN-IBM %\cite{desai_graph_2021}
\\
\hline    SM & 0.28 & 0.25 & 0.42 & \textbf{0.29} & 0.28 %& 0.29 
\\
\hline    IFN &  1.3 & 1.5 & 0.8 & 1.7 & \textbf{0.8} %& 3 
\\
\hline    NED & 0.7 & \textbf{0.3} & 1 & 0.7 & 0.35 %& \textbf{0.1} 
\\
\hline    ICP &  0.5 & 0.5 & 0.62 & 0.72 & \textbf{0.1}% &  
\\

\hline
\end{tabular}
 \caption{Summary table of evaluation results of state of the art approaches for the
Acmeair application}
  \label{Summary table of evaluation results of state of the art approaches for the Acmeair application}
\end{table}

Following the detected values for the different approaches presented in Table 5.1, we
can notice that the CoGCN approach determined in the paper "HierDecomp"  \cite{sellami_hierarchical_2022} has the
value NED=1 which is an extreme result. This proves that all microservices formed by
this approach are very small or very large. Therefore, we decide to eliminate it from the
comparison approaches because it does not respect one of the main objectives of the
migration. This famous problem is known in literature under the name of the "Problem of grains and borders"

%tableau DayTrader littérature
\begin{table}[htbp]
\centering
\begin{tabular}{|r|m{3cm}|m{2cm}|m{2cm}|m{2cm}|m{2cm}|}
\hline  DayTrader & Mono2Micro-IBM %\cite{kalia_mono2micro_2021}
& Bunch-HierDecomp %\cite{sellami_hierarchical_2022} 
& GNN %\cite{sellami_hierarchical_2022}
& FoSCI-HierDecomp %\cite{sellami_hierarchical_2022}
& MEM-HierDecomp  %\cite{sellami_hierarchical_2022}
%& GNN-IBM %\cite{desai_graph_2021}
\\
\hline    SM & \textbf{0.53}& 0.15& 0.5& 0.29 & 0.3% & 0.2 
\\
\hline    IFN & 3& \textbf{0.1}& 13 & 6& 3.5% & 3
\\
\hline    NED & \textbf{0.3} & 0.65 & 0.65 & 0.58& 1% & \textbf{0.15}
\\
\hline    ICP & \textbf{0.1} & 0.5 & 0.3 & 0.8 & 0.25 
\\

\hline
\end{tabular}
 \caption{ Summary table of the results of the evaluation of the state of the art approaches for the DayTrader application}
  \label{ Summary table of the results of t h e evaluation of the s t a t e - o f-t h e - a r t approaches for the DayTrader application}
\end{table}
For the same reasons cited above, we eliminate the MEM approach from table 5.2.
\subsection{Results of the naive approach and the codependent calls approach}
\paragraph{}
The following table summarizes the results obtained for our first two approaches using
both DBSCAN and BMSC algorithms for each.

\begin{table}[htbp]
\centering
\begin{tabular}{|r|m{3cm}|m{2cm}|m{2cm}|m{2cm}|m{2cm}|}
\hline \textbf{Acmeair}  & Algorithms & Naive
approach & Co-dependent call approach \\
\hline   \multirow{2}*{SM} & \multirow{1}*{DBSCAN} & 0.16 & 0.17 \\

\cline{2-4}  & \multirow{1}*{BMSC} & 0.42 & \textbf{0.44}   \\

\hline   \multirow{2}*{IFN} & \multirow{1}*{DBSCAN} & 0.5 & 1 \\

\cline{2-4}  & \multirow{1}*{BMSC} & 4.8 & \textbf{0.2}  \\

\hline   \multirow{2}*{NED} & \multirow{1}*{DBSCAN} & 0.8 & 1 \\

\cline{2-4}  & \multirow{1}*{BMSC} & 0.8 & \textbf{0.8}  \\

\hline   \multirow{2}*{ICP} & \multirow{1}*{DBSCAN} & 0.48 & 0.56  \\

\cline{2-4}  & \multirow{1}*{BMSC} & 0.58 & \textbf{0.4}   \\

\hline
\end{tabular}
 \caption{ Summary table of the results of the first two approaches - Acmeair }
  \label{- Summary table of the results of the first two approaches - Acmeair}
\end{table}
As we can notice from Table 5.3 of the Acmeair application, the best result for this
application was generated by the second approach of codependent calls.
\newline On the other hand, for this same approach of codependent calls, BMSC has shown to be
more efficient than DBSCAN and this by comparing the values obtained for the different
evaluation metrics. Indeed, the best result for SM is 0.44 resulting from the application of
BMSC for the second resolution approach. The same is true for all the other metrics.
\paragraph{}
Since these evaluations, we find that the second approach is more efficient than the
naive one, due to the refinement of the class representation criterion as explained in the
third chapter of our report. 
\newline A further evaluation shows that BMSC performs better than DBSCAN when the data
have a wide range of density. This confirms the results shown by Ren et al \cite{calders_boosted_2014}.\\
In a second phase, we compare our results with those of the literature presented in
table 5.1.
\newline Our solution proves the best result for SM and NED metrics with 0.44 and
0.2 respectively compared to the best values found which are SM=0.42 for CoGCN and
IFN=0.8 for this same approach.

\begin{table}[htbp]
\centering
\begin{tabular}{|r|m{3cm}|m{3cm}|m{3cm}|m{3cm}|m{3cm}|}
\hline \textbf{DayTrader}  & Algorithms & Naive approach  & Co-dependent calls approach\\
\hline   \multirow{2}*{SM} & \multirow{1}*{DBSCAN} & 0.42 & 0.2 \\

\cline{2-4}  & \multirow{1}*{BMSC} & \textbf{0.7} & \textbf{0.7}   \\

\hline   \multirow{2}*{IFN} & \multirow{1}*{DBSCAN} & 0 & 2.2 \\

\cline{2-4}  & \multirow{1}*{BMSC} & 11& \textbf{1.6} \\

\hline   \multirow{2}*{NED} & \multirow{1}*{DBSCAN} & 1 & 0.8 \\

\cline{2-4}  & \multirow{1}*{BMSC} & 0.85 & \textbf{0.8}  \\

\hline   \multirow{2}*{ICP} & \multirow{1}*{DBSCAN} & 0.57 & 0.52  \\

\cline{2-4}  & \multirow{1}*{BMSC} & 0.54 & \textbf{0.5}   \\

\hline
\end{tabular}
 \caption{ Summary table of the results of the first two approaches - DayTrader }
  \label{- Summary table of the results of the first two approaches - DayTrader}
\end{table}

DayTrader also confirms through the results generated the performance of the second
approach for the application of the BMSC algorithm compared to the first naive and
DBSCAN.
\newline As shown in Table 5.4 of the DayTrader application, we obtained the same best SM value
for 0.7 when applying BMSC for both approaches. This puts both approaches on the same
line of efficiency but if we move forward in our evaluation by exploiting the other
aspects of the newly trained system, we notice that the preferred results are those of the second approach with NFI = 1.6 against NFI=11
during the first approach. Similarly, NED having its best value for 0.8 and finally ICP for
the value of 0.5.
\paragraph{}
Also, DayTrader achieves the best results for the SM, IFN metrics compared to all
approaches except the Bunch method.
\subsection{Results of the graph approach}
\paragraph{}
The following table summarizes the results obtained for our last reso- lution approach
based on graph theory. 

\begin{table}[htbp]
\centering
\begin{tabular}{|r|m{3cm}|m{3cm}|m{3cm}|m{3cm}|m{3cm}|}
\hline \textbf{Metrics}  & Algorithms & \textbf{Acmeair}  & \textbf{DayTrader} \\
\hline   \multirow{2}*{SM} & \multirow{1}*{Girvan Newman} & 0.7 & 0.3 \\

\cline{2-4}  & \multirow{1}*{Louvain} & 0.5 & 0.4   \\

\hline   \multirow{2}*{IFN} & \multirow{1}*{Girvan Newman} & 1 & 0.6 \\

\cline{2-4}  & \multirow{1}*{Louvain} & 1.4& 0.8 \\

\hline   \multirow{2}*{NED} & \multirow{1}*{Girvan Newman} & 1 & 0.8 \\

\cline{2-4}  & \multirow{1}*{Louvain} & 0.3 & 0.6  \\

\hline   \multirow{2}*{ICP} & \multirow{1}*{Girvan Newman} & 0.1 & 0.52 \\

\cline{2-4}  & \multirow{1}*{Louvain} & 0.6 & 0.85   \\

\hline
\end{tabular}
 \caption{ Summary table of the results of the graph approach }
  \label{- Summary table of the results of the graph approach}
\end{table}
%\vspace{-1cm}
From table 5.5, we notice that for the Acmeair application, the Girvan-Newman
algorithm generates the best results. The value of NED being extreme equal to 1 pushes us
to further detail the evaluation and this through figure 5.1.\\
\newline Note that the nodes with the same color in the figure belong to the same cluster, in other
words the classes sharing the same color of nodes belong to the same microservice. The
figure shows that the Girvan-Newman algorithm and to maximize the value of structural
modularity tried to gather all the classes in the same microservice except two that were
each assigned to a microservice. This resulted in the formation of one very large
microservice and two very small microservices which does not meet the NED metric.\\
On the other hand and for the same Acmeair application, the Louvain community
detection algorithm generated four medium-sized clusters as shown in Figure
5.2 thus justifies the value obtained for the NED metric which is equal to 0.4.

\begin{figure}[htbp]
    \centering
    \includegraphics [bb=180 0 5cm 6cm]{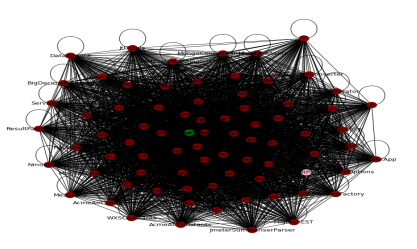}
    \caption{Community graph of the application Acmeair using Girvan Newman algorithm
 }
    \label{Fig:Community graph of the application Acmeair using Girvan Newman algorithm  }
\end{figure}

\begin{figure}[htbp]
    \centering
    \includegraphics[bb=180 0 5cm 8cm]{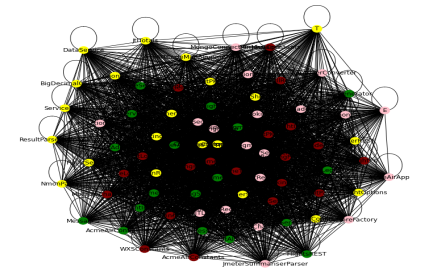}
    \caption{Community graph of the application Acmeair using  Louvain algorithm }
    \label{Fig:Community graph of the application Acmeair using  Louvain algorithm}
\end{figure}

A comparative study between the results of approach 2 and approach 3 proves that the
codependent calls approach tends to generate the best results in terms of structural
modularity but the graph approach focuses much more on minimizing the metrics
considering even more the coupling.
\subsection{Discussion}
\paragraph{}
Comparing to the reference approaches, our solution dealing with the problem from a
clustering point of view using the BMSC algorithm outperforms all other solutions in the
literature in terms of structural modularity for both Acmeair and DayTrader test
applications. Nevertheless, it is worse than some other methods for the rest of the metrics.
\paragraph{}
This problem of unsatisfiability of all evaluation metrics is manifested in all
approaches as we note from Tables 5.1 and 5.2.
\newline No approach in the literature presents the best performances for all the proposed
evaluation metrics but to be able to conclude about the reliability of the approaches and
according to the literature an approach is classified as acceptable if it reaches good results
for the structural modularity provided that it does not contain extreme values for NED and
IFN.\\
Therefore, we can conclude that our second approach based on the consideration of
codependent calls by applying the BMSC algorithm satisfies the resolution of the migration problem by presenting performances that exceed those of the literature approaches presented in Tables 5.1 and 5.2.\vspace{2cm}

\textbf{\LARGE Conclusion}
\paragraph{}
During this last chapter of our report, we started by presenting our working
environment with these two software and hardware parts. Then, we summarized the results
of different approaches in the literature and presented our own in a second part. Finally,
we concluded the chapter with a discussion of the results.

%% file: Chapitres/conclusion.tex
\chapter*{Conclusion and perspectives}

\label{sec:conclusion}

\paragraph{}
In this report, we present an approach to decompose monolithic applications into microservices based on a static analysis of the source code of the latter while treating the
problem as a clustering problem.

We began by introducing the general context of the project to lay the
groundwork. Then, we have introduced the different key concepts necessary for a good
understanding of the theme. We also reviewed existing approaches using a well-defined
research methodology. This review was the key to establishing our approach. It is
conducted to inquire about the main results of existing similar works and their limitations,
which helped us to define the contributions of the work and to make judicious choices on
the methodology of treatment of the problem. Finally, the report details the work done
starting with the understanding, collection and processing of the data up to the modeling
part. We conclude our report with the results of our approach and a comparison with
existing approaches.
\paragraph{}
At the end of this course, we can conclude that we have obtained good results since we
have outperformed existing solutions in many evaluation metrics. The best results we
have detected are those generated by the second approach using codependent calls
between classes of the application precisely by applying the BMSC clustering algorithm.
\paragraph{}
During this project, we encountered many problems when establishing the
decomposition approach such as the definition of the decomposition criteria and the level
of granularity of the solution as well as the method and metrics to be used for the
evaluation of the results, from which the need for the research was born. 
\paragraph{}
Therefore, we could continue this project by: further developing a complement to the
evaluation strategy of our project that corresponds to a semi-automatic decomposition
process. This process will be considered as the basis for per-application comparison since
the evaluation metrics used in the literature do not cover all aspects of the monolithic
application decomposition.
\newline In addition, we can continue to exploit graph theory by testing the notion of deep
community detection.
\newline Moreover, the choice of the granularity level is a decisive step of the project, so we can
push the solution even further to not consider classes as a basis for decomposition but
rather methods or functions of the monolith.
\newline Also, we can make our solution hybrid by injecting a dynamic analysis of the source code
based on the fact that the static analysis does not provide all the elements necessary for the
good understanding of the functionalities and their interactions during the execution of the
application.
\newline A final perspective is to not only recommend a way to decompose the monolith but also to
recommend a way to rewrite the application code if necessary.\\